\documentclass[journal=jpcbfk]{achemso}
\setkeys{acs}{articletitle=true}
\usepackage[version=3]{mhchem}
\usepackage{graphicx}
\usepackage{amsmath}
\usepackage{amssymb} 
\usepackage{mhchem}

\author{Alexandre P. dos Santos}
\email{alexandre.pereira@ufrgs.br}
\affiliation{Instituto de F\'isica, Universidade Federal do Rio Grande do Sul, Caixa Postal 15051, CEP 91501-970, Porto Alegre, RS, Brazil}

\author{Matheus Girotto}
\email{matheus.girotto@ufrgs.br}
\affiliation{Instituto de F\'isica, Universidade Federal do Rio Grande do Sul, Caixa Postal 15051, CEP 91501-970, Porto Alegre, RS, Brazil}

\author{Yan Levin}
\email{levin@if.ufrgs.br}
\affiliation{Instituto de F\'isica, Universidade Federal do Rio Grande do Sul, Caixa Postal 15051, CEP 91501-970, Porto Alegre, RS, Brazil}
\phone{55 51 33086517}

\title{Simulations of Polyelectrolyte Adsorption to a Like-Charged Membrane}

\begin{document}

\begin{abstract}

We explore, using recently developed efficient Monte Carlo simulation method,  
interaction of anionic polyelectrolyte solution with a like-charged 
membrane.  In addition to polyions, solution also contains salt with either monovalent,  
divalent, or trivalent counterions. In agreement with recent experimental observations, we find that multivalent counterions can lead to strong adsorption of polyions to a like charged surface.  On the other hand, addition of 1:1 electrolyte diminishes the adsorption induced by the multivalent counterions. Dielectric discontinuity at the interface is found to play only a marginal role for the polyion adsorption.   

\end{abstract}

\maketitle

\section{Introduction}

Charged polymers are very important for different areas. They are used in industry as purifying agents~\cite{BaPa02,WaJa13}. The strong electrostatic interactions promoted by polyelectrolytes can induce phase transitions in a solution of bigger oppositely charged objects. This can lead to coagulation, separating undesirable species from water. Polyelectrolytes are used in medical applications as antibiotics~\cite{CaUl04,LuRi05,KiHe12,HoAk12,FaFa12}, and in diagnostics as chemical sensors~\cite{FeLi10,KiGu11}. For amphiphilic polyelectrolytes the antimicrobial behavior appear to be related with the electrostatic interactions between the negative charges of the membrane and the positive charges of the polyion, as well as the hydrophobic interactions between the phospholipids and the polyelectrolytes. In our body, DNA is a negatively charged polyion which can interact with positively charged liposomes or proteins. A number of groups have studied charged polymers interacting with oppositely charged surfaces~\cite{KoMu98,ChSt01,DoDe00,DoDe01,DoRu02,McKo02,ReYe10,QiCe11,FaMa13}. Some groups focused on curvature effects, studying the adsorption on a oppositely charged spherical surface~\cite{StCh02,MeHo04}. The Hofmeister effect, or ionic specificity, was observed in the adsorption of polyions to interfaces~\cite{SaKa08,DoLe13,KoZh15} besides being observed in a lot of other systems. Other authors have investigated the attraction between like-charged macroions~\cite{HaLi99,ArSt99,DeJi01,BuAn03,DeAr03,AnLi03,MoRi05}. The attraction is short ranged and depends on the presence of multivalent counterions. The mechanism responsible are the correlations between condensed counterions surrounding the polyelectrolytes. However, only recently some groups have started to explore the interaction between polyelectrolyte and like-charged membranes~\cite{TuLa11,LuMa14,TiMa15}. This study can lead to important applications in future.

Simulations of Coulomb systems are difficult because of the long range nature of the Coulomb force~\cite{AlTi87}. In a slab geometry the complexity increases further because of the reduced symmetry which complicates implementation of the periodic boundary conditions using standard Ewald summation techniques. Different methods have been devised in order to overcome these difficulties~\cite{ToVa80,ToVa84,Le91,WiAd97,YeBe99,ArHo02}. Recently, we introduced a new method which allows us to efficiently perform simulations of systems with underlying slab geometry~\cite{DoGi16}. The method was developed specifically to study inhomogeneous Coulomb systems near charged surfaces, such as electrodes, membranes, or large colloidal particles.  The idea of the new approach is to separate the electrostatic potential produced by the uniformly charged surface from the other electrostatic interactions, threating it as an external potential acting on ions and polyions. The difficulty, however, is that such separation results in a non-neutral Coulomb system which, when treated using regular Ewald summation, leads to infinite electrostatic self energy.  However, we were able to show that this infinite contribution can be renormalized away, resulting in a well defined finite electrostatic energy which can be used within the Metropolis algorithm to very efficiently perform Monte Carlo simulations. A similar approach was also recently proposed~\cite{BaAr09,Ba14}, although the derivations and motivations were different from ours. Their motivation was the study of slab systems with a net charge. In the present paper we will show how to modify the energy expressions derived in Ref.~\cite{DoGi16} in order to study systems with dielectric discontinuities~\cite{DoLe14b}.

A particularly interesting application of the new algorithm, presented in the present paper, is to explore adsorption of polyions to a like-charged dielectric surface inside an electrolyte solution containing multivalent and monovalent counterions. In this paper we will show how these systems can be efficiently simulated using the new algorithm based on non-neutral Ewald 3d summation method. In particular we will explore the effect of  dielectric surface polarization on the interaction between anionic polyelectrolyte and a like-charged membrane. In the following sections we will present the computational details, results, discussions, and the conclusions of the present study. All the technical details of the derivations will be provided in the appendix.

\section{Computational Details}

Our system consists of a negatively charged membrane with a surface charge density $\sigma$, anionic polyelectrolyte, 
and dissolved salt. The simulation box has sides $L_x=160.1~$\AA, $L_y=L_x$ and $L_z=4 L_x$. The electrolyte is confined in $-L_x/2 < x < L_x/2$, $-L_y/2 < y < L_y/2$ and $0 < z < L$, where $L=250~$\AA, while empty space is maintained in the complementary region. The charged surface is located at $z=0$, while the confining neutral surface is located at $z=L$. The primitive model is considered. The polyions are modeled as flexible linear chains of $N_m$ spherical monomers of charge $-q$ adjacent to each other, where $q$ is the proton charge. Besides ions from $\alpha:1$ salt at concentration $\rho_S$, where $\alpha$ is the cationic valence, additional monovalent counterions of charge $q$ that neutralize polyion and surface charges are also present. The effective diameter of all ions and monomers is set to $4~$\AA. The water is a continuum medium of dielectric constant $\epsilon_w=80\epsilon_0$, where $\epsilon_0$ is the dielectric constant of the vacuum. The Bjerrum length, defined as $\lambda_B=q^2/\epsilon_wk_BT$, is set to $7.2~$\AA, value for water at room temperature. In the region of empty space, $z<0$, we consider that the medium is a continuum with dielectric constant $\epsilon_m$, modeling the membrane. The total energy of the system is given by
\begin{equation}
U=U_S+U_{self}+U_L+U_{cor}+U_P+U_{pol} \ .
\end{equation}
In the present method, the details of which are presented in appendix, 
the electrostatic interaction energy between an {\it infinite} charged dielectric surface and all 
the charged particles is given by,
\begin{equation}
U_P=-\frac{2\pi}{\epsilon_w}(1+\gamma)\sum_{i=1}^N q_i z_i \sigma \ ,
\end{equation}
where $\gamma=(\epsilon_w-\epsilon_m)/(\epsilon_w+\epsilon_m)$. The rest of the polyelectrolyte-electrolyte system, {\it which is not charge neutral}, is treated using periodic 3d Ewald summation.
The Coulomb potential is split into short range and long range contributions.  The short range 
part can be studied using simple periodic boundary conditions, while the long-range contribution
can be efficiently evaluated in the reciprocal Fourier space.
The short range electrostatic energy is 
\begin{equation}
U_S=(1/2)\sum_{i=1}^{N}q_i\phi_i^S({\pmb r}_i) \ ,
\end{equation}
where $\phi_i^S({\pmb r})$ is,
\begin{equation}\label{phi_S3}
\phi_i^S({\pmb r})=\sum_{j=1}^{N}{}^{'} q_j\frac{\text{erfc}{(\kappa_e |{\pmb
r}-{\pmb r}_j|)}}{\epsilon_w |{\pmb r}-{\pmb r}_j|} + \sum_{j=1}^{N}\gamma q_j
\frac{\text{erfc}{(\kappa_e |{\pmb r}-{\pmb r}'_j|)}}{\epsilon_w |{\pmb r}-{\pmb
r}'_j|} \ ,
\end{equation}
where $N$ refers to all the charges in the simulation box, except the  wall, ${\pmb r}_j$ is the position of the charge $q_j$ and ${\pmb r}'_j={\pmb r}_j-2z_j \hat{\pmb z}$ is the position of the image charge $\gamma q_j$. The prime on the summation means that $j\neq i$. The damping parameter is set to $\kappa_e=4/L_x$. The self-energy contribution is
\begin{equation}
U_{self}=-(\kappa_e/\epsilon_w\sqrt{\pi})\sum_{i=1}^{N}q_i^2 \ .
\end{equation}

The long range electrostatic energy is
\begin{eqnarray}\label{U_long}
U_L = \sum_{{\pmb k}\neq {\pmb 0}}\frac{2\pi}{\epsilon_w V |{\pmb k}|^2}
\text{exp}(-\frac{|{\pmb k}|^2}{4\kappa_e^2}) \times \nonumber \\
\left[A({\pmb k})^2 + B({\pmb k})^2 + A({\pmb k}) C({\pmb k}) + B({\pmb k}) D({\pmb k})\right] \ ,
\end{eqnarray}
where
\begin{eqnarray}
A({\pmb k})= \sum_{i=1}^{N}q_i \cos{({\pmb k}\cdot {\pmb r}_i)} \ , \nonumber \\
B({\pmb k})=- \sum_{i=1}^{N}q_i \sin{({\pmb k}\cdot {\pmb r}_i)} \ , \nonumber \\
C({\pmb k})= \sum_{i=1}^{N}\gamma q_i \cos{({\pmb k}\cdot {\pmb r}'_i)} \ , \nonumber \\
D({\pmb k})=- \sum_{i=1}^{N} \gamma q_i \sin{({\pmb k}\cdot {\pmb r}'_i)} \nonumber \ .
\end{eqnarray}
The number of vectors ${\pmb k}$, defined as ${\pmb k}=(2\pi n_x/L_x,2\pi n_y/L_y,2\pi n_z/L_z)$, where $n's$ are integers, is set to around $400$ in order to achieve convergence.

The correction, which accounts for the conditional convergence of the Ewald summation  appropriate for 
the slab geometry, is derived in the appendix. It is given by
\begin{equation}\label{ucor}
U_{cor}=\frac{2\pi}{\epsilon_w V} \left[ M_z^2 (1-\gamma) - G_z Q_t (1+\gamma) \right] \ ,
\end{equation}
where
\begin{equation}
M_z=\sum_{i=1}^{N}q_i z_i \ \ \text{,} \ \ G_z=\sum_{i=1}^{N}q_i z_i^2 \ \ \text{and} \ \ Q_t=\sum_{i=1}^{N}q_i \ .
\end{equation}
Note that this correction depends on the net charge $Q_t$ present inside the simulation cell, without
including the surface charge of the membrane, which has already been accounted for in $U_P$.

The monomers that compose a polyion interact via Coulomb potential and via a simple parabolic potential~\cite{DiPa05,QuMa13,LuMa14} which models stretching of molecular bonds,
\begin{equation}
U_{pol}=\sum_{ad. mon.} \dfrac{A}{2}(r-r_0)^2 \ .
\end{equation}
The sum is over the adjacent monomers, where $r$ is the distance between the adjacent monomers, and $A=0.97~k_BT$, and $r_0=5~$\AA. 

The simulations are performed using Metropolis algorithm with $10^{6}$ MC steps for equilibration. The profiles are calculated with $5\times 10^4$ uncorrelated states, each obtained with $100$ trial movements per particle. Polyions can perform rotations and reptation moves.  In addition, polyion monomers can attempt short displacements, while ions can perform both short and long distance moves.

\section{Results \& Discussion}

\begin{figure}[h]
\vspace{0.2cm}
\includegraphics[width=7cm]{fig1.eps}\vspace{0.2cm}
\caption{Comparison between PCM concentration versus distance from surface, obtained with $\epsilon_m=\epsilon_0$, and with $\epsilon_m=\epsilon_w$, corresponding to absence of membrane polarization. The membrane charge density is $\sigma=-0.1~$C$/$m$^2$, the number of polyions in simulation box is $20$, $N_m=18$, $\alpha=3$, and salt concentration is $60~$mM.}
\label{fig1}
\vspace{0.2cm}
\includegraphics[width=7cm]{fig2.eps}\vspace{0.2cm}
\caption{Comparison between trivalent cationic concentration versus distance from surface, obtained with and without membrane polarization. The parameters are the same as in Fig.~\ref{fig1}.}
\label{fig2}
\end{figure}
\begin{figure}[h]
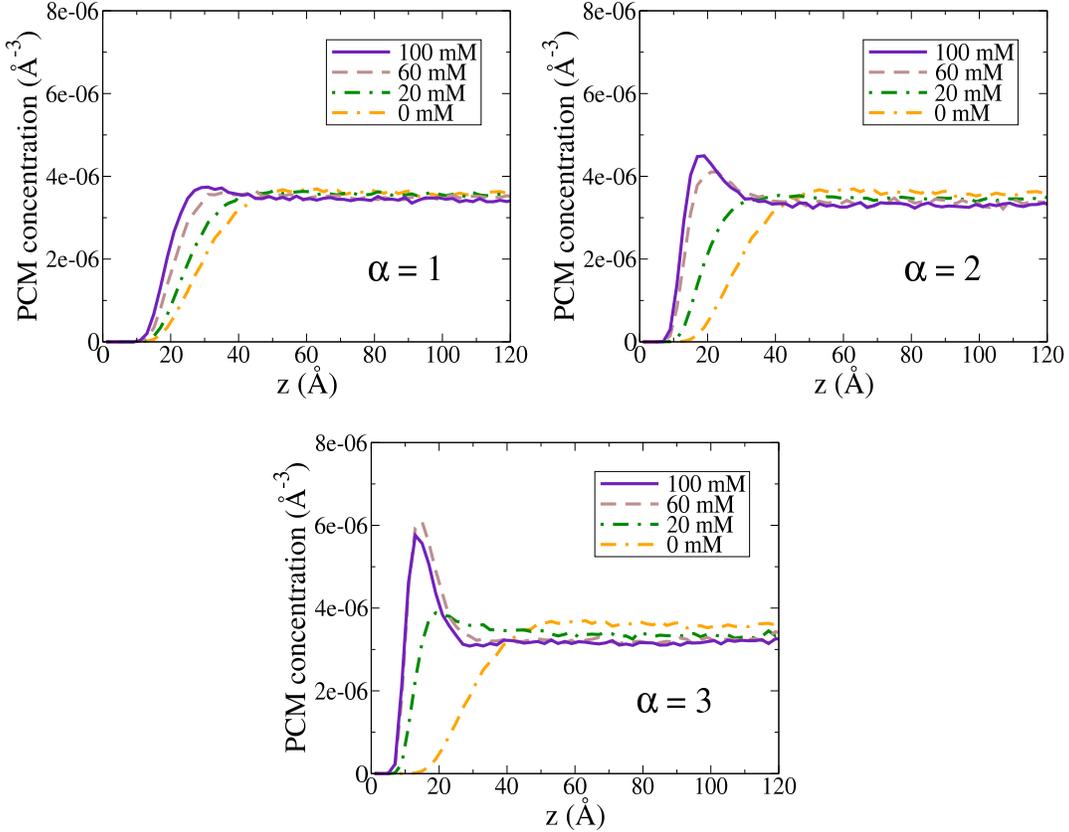

\vspace{0.2cm}
\includegraphics[width=7.cm]{fig3a.eps}\vspace{0.2cm}
\includegraphics[width=7.cm]{fig3b.eps}\vspace{0.2cm}
\includegraphics[width=7.cm]{fig3c.eps}\vspace{0.2cm}
\caption{Comparison between PCM concentration versus distance from surface obtained for different salt concentrations. The parameters are the same as in Fig.~\ref{fig1}, except $\alpha$ and salt concentrations. The valence of $\alpha:1$ salt is $\alpha=1$, $2$ and $3$.}
\label{fig3}
\end{figure}
First, we explore the effect of the dielectric discontinuity on the interaction of polyelectrolyte with  membrane. We consider two cases: when $\epsilon_m=\epsilon_w$ and when $\epsilon_m=\epsilon_0$. 
The effect of surface polarization on ionic double layer was extensively studied for different geometries~\cite{KaNa10,NaHe11,SaTr11,BaDo11,DoBa11,DiDo12,LuLi11,WaWa13,JiJa15,DoLe15}. In Fig.~\ref{fig1}, the polyion center of mass~(PCM) distribution is shown as a function of the distance from the wall, for a system with $\alpha=3$. We see that the effect of membrane polarization is surprisingly small. Although the 
multivalent counterions are strongly repelled from the surface, see Fig.~\ref{fig2}, the net effect on the polyion adsorption is minimal, with only a slight change in the equilibrium position of the polyion center of mass distribution. This effect is related with the dense structure of a polyion and the local behavior of multivalent ions. The polyelectrolyte structure leads to peak to stay around $12~$\AA, see Fig.~\ref{fig1}, for example. This distance is far from the peak of trivalent ions for same parameters, around $4~$\AA, see Fig.~\ref{fig2}.
For all the following results, we will take into account membrane polarization by setting $\epsilon_m=\epsilon_0$.

\begin{figure}[t]
\vspace{0.2cm}
\includegraphics[width=7.cm]{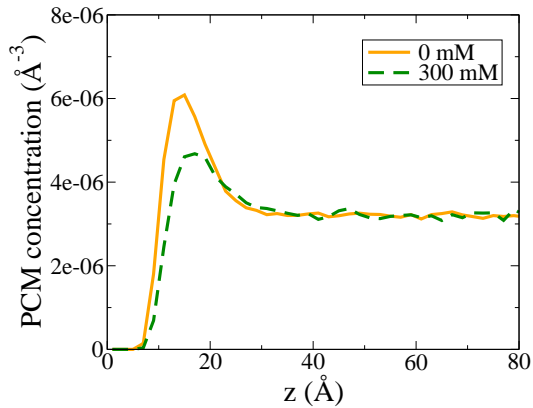}\vspace{0.2cm}
\caption{Comparison between PCM density versus distance from surface obtained with and without addition of 1:1 salt. The concentration of 3:1 electrolyte is fixed at $60~$mM. The parameters are the same as in Fig.~\ref{fig1}.}
\label{fig4}
\end{figure}

The effect of salt concentration on adsorption is shown in Fig.~\ref{fig3}, where the results are presented for $\alpha=1$,  $\alpha=2$, and $\alpha=3$. In the absence of salt, as can be seen, the polyions do not adsorb to a like-charged membrane. Increasing concentration of $\alpha=1$ electrolyte enhances the electrostatic screening decreasing the repulsion, leading to a small adsorption for high salt, in agreement with the recent experimental results~\cite{TiMa15}. For $\alpha=2$ electrolyte, there is a significant adsorption of polyions onto a like-charged membrane, as was also observed in experiments~\cite{TiMa15}. We next consider $\alpha=3$ electrolyte. In this case the  adsortion of polyions is very strong, mediated by the positional correlations of the condensed multivalent counterions, in agreement with the previous simulation results~\cite{LuMa14}. We also see that the adsorption decreases for sufficiently large concentrations of $\alpha=3$ electrolyte, which is again consistent with the  experimental observations~\cite{TiMa15}. This behavior is related with the saturation of condensed trivalent ions and increasing of electrostatic screening with added salt, similar to observations of simulations with added divalent salt~\cite{TuLa11}. With a sufficiently high salt concentration the adsorption starts to decrease also for monovalent and divalent salts, as shown in experiments~\cite{TiMa15} and simulations~\cite{TuLa11}, however we do not achieve such high concentrations for these salts in this work. Finally, in Fig.~\ref{fig4} we explore the effect of addition of 1:1 electrolyte to polyelectrolyte solution containing $\alpha=3$ electrolyte at $60~$mM concentration.  We see that addition of 1:1 electrolyte screens the electrostatic interactions, diminishing the adsorption of polyions to a like-charged surface~\cite{PiBa05}.   

The mechanism responsible for the polyion attraction to a like-charged membrane was partially attributed to hydrophobic interactions between the charged polymers and membrane, for moderate trivalent salt concentrations~\cite{TiMa15}. However in this paper, we see that purely electrostatic interactions, without any specific hydrophobic effects, already can result in a like-charge attraction. It was also suggested~\cite{TiMa15} that the charge inversion of the polyion-cation complex is responsible for the polyion condensation onto a like-charged surface. However, our simulations show that, in general, this is not the case. In the  presented model, the polyions are not sufficiently charged to result in a charge reversal of the polyion-cation complex. In simulations we see that the attraction is a consequence of strong electrostatic correlations between adsorbed multivalent ions~\cite{Sh99,DoDi09,Le02} and the polyion monomers.  Electrostatic correlations have been previous found to be also responsible for the inversion of electrophoretic mobility~\cite{QuCa02,FeFe05} and attraction between like-charged colloidal particles~\cite{SoDe01,Le02}.


\section{Conclusions}

We have presented MC simulations of polyelectrolyte solutions interacting with 
like-charged membranes.  The simulations
were performed using a recently developed algorithm, which allows us to efficiently study inhomogeneous Coulomb systems with a planar charged interface~\cite{DoGi16}.
The effect of membrane polarization, which results in induced surface charge,  has been taken into account using image charges. 
The adsorption has been characterized by the PCM distribution. Surprisingly, we find a small adsorption of polyions onto a like-charged membrane even for $\alpha=1$ electrolyte at sufficiently large concentrations.
In this case, the electrostatic correlations do not play any significant role and the attraction is a consequence of steric and depletion interactions. For $\alpha=2$ electrolyte, for all the conditions studied in the paper, adsorption increased with salt concentration. For $\alpha=3$ electrolyte, the polyion adsorption first increased with the concentration of multivalent salt, and then decreased. Addition of 1:1 electrolyte to a polyelectrolyte solution containing multivalent counterions decreased the polyion adsorption. All the results are consistent with the recent experimental observations.  

\section{Acknowledgments}
This work was partially supported by the CNPq, INCT-FCx, and by the US-AFOSR under the grant FA9550-12-1-0438.

\appendix

\section{Energy Calculations}

We consider a system of $N$ charged particles with charges $q_j$ located at ${\pmb r}_j$ bounded by a dielectric wall at $z=0$. The simulation box has sides $L_x$, $L_y$ and $L_z$ and volume $V=L_xL_yL_z$. The electrolyte is confined in $-L_x/2<x<L_x/2$, $-L_y/2<y<L_y/2$ and $0<z<L$. The system is in general not charge neutral. In order to take into account the long range nature of the Coulomb force, we replicate the system periodically in all directions.  The ions in the main simulation cell interact with all the other ions in the cell and also with all the periodic replicas.  We define the replication vector as ${\pmb r}_{rep}=(n_xL_x,n_yL_y,n_zL_z)$, where $n$'s are integers. To correctly simulate the system we have to consider the polarization of the dielectric wall, which can be done by  introducing image charges. The potential due to the real and image charges at an arbitrary position ${\pmb r}$ inside the main simulation cell is
\begin{eqnarray}\label{elec_pot}
\phi_i({\pmb r})=\sum_{\pmb n}^{\infty}{}^{'}\sum_{j=1}^{N}\int\frac{\rho_j({\pmb s})}{\epsilon_w |{\pmb r}-{\pmb s}|}d^3{\pmb s} + \nonumber \\
\sum_{\pmb n}^{\infty}\sum_{j=1}^{N}\int\frac{\rho_j'({\pmb s})}{\epsilon_w |{\pmb r}-{\pmb s}|}d^3{\pmb s} \ ,
\end{eqnarray}
where $\rho_j({\pmb s})=\text{q}_j \delta({\pmb s}-{\pmb r}_j-{\pmb r}_{rep})$ is the
charge density of ion $j$ and its infinite replicas, and $\rho_j'({\pmb s})=\gamma\text{q}_j \delta({\pmb s}-{\pmb r}_j'-{\pmb r}_{rep})$ is the image charge density of ion $j$ and its infinite replicas. The prime over summation means that $i\neq j$ for ${\pmb n=(0,0,0)}$. The constant $\gamma$ assumes the value $\gamma=(\epsilon_w-\epsilon_m)/(\epsilon_w+\epsilon_m)$, where $\epsilon_m$ is the dielectric constant of the surface medium and $\epsilon_w$ is the dielectric constant of the medium where the real charges are. The vector ${\pmb r}_j'$ is the position of the image charges defined as ${\pmb r}_j'={\pmb r}_j-2z_j \hat{\pmb z}$. The vectors ${\pmb n}=(n_x,n_y,n_z)$ represent the different replicas.

We use 3d Ewald summation technique to efficiently sum over the replicas. The potential has the form
\begin{eqnarray}\label{elec_pot_E}
\phi_i({\pmb r})=\sum_{\pmb n}^{\infty}{}^{'}\sum_{j=1}^{N}\int\frac{\rho_j({\pmb s})-\rho_j^G({\pmb s})}{\epsilon_w |{\pmb r}-{\pmb s}|}d^3{\pmb s} + \nonumber \\
\sum_{\pmb n}^{\infty}\sum_{j=1}^{N}\int\frac{\rho_j'({\pmb s})-\rho_j'^G({\pmb s})}{\epsilon_w |{\pmb r}-{\pmb s}|}d^3{\pmb s} + \nonumber \\
\sum_{\pmb n}^{\infty}\sum_{j=1}^{N}\int\frac{\rho_j^G({\pmb s})}{\epsilon_w |{\pmb r}-{\pmb s}|}d^3{\pmb s} + \nonumber \\
\sum_{\pmb n}^{\infty}\sum_{j=1}^{N}\int\frac{\rho_j'^G ({\pmb s})}{\epsilon_w |{\pmb r}-{\pmb s}|}d^3{\pmb s} - \nonumber \\
\int\frac{\rho_i^G({\pmb s})}{\epsilon_w |{\pmb r}-{\pmb s}|}d^3{\pmb s} \ ,
\end{eqnarray}
where
\begin{eqnarray}\label{rhoG}
\rho_j^G({\pmb s})=\text{q}_j (\kappa_e^3/\sqrt{\pi^3})\exp{(-\kappa_e^2|{\pmb s}-{\pmb r}_j-{\pmb r}_{rep}|^2)} ,\nonumber \\
\rho_j'^G({\pmb s})=\gamma\text{q}_j (\kappa_e^3/\sqrt{\pi^3})\exp{(-\kappa_e^2|{\pmb s}-{\pmb r}_j'-{\pmb r}_{rep}|^2)} \ ,
\end{eqnarray}
and $\kappa_e$ is a damping parameter. The first two terms of Eq.~\ref{elec_pot_E} define a short range potential, $\phi_i^S({\pmb r})$,
\begin{eqnarray}\label{phis}
\phi_i^S({\pmb r})=\sum_{j=1}^{N}{}^{'}\text{q}_j\frac{\text{erfc}(\kappa_e|{\pmb r}-{\pmb r}_j|)}{\epsilon_w |{\pmb r}-{\pmb r}_j|} + \nonumber \\
\sum_{j=1}^{N}\gamma\text{q}_j\frac{\text{erfc}(\kappa_e|{\pmb r}-{\pmb r}_j'|)}{\epsilon_w |{\pmb r}-{\pmb r}_j'|} \ .
\end{eqnarray}
Notice that we can exclude the summation over $\pmb n$'s in the short range potential, adopting the usual minimum image convention, $\pmb n=(0,0,0)$. This is appropriate because of the exponentially fast decay of $\text{erfc}(x)$ with increasing $x$. The total short range interaction energy is then
\begin{equation}
U_S=(1/2)\sum_{i=1}^{N}q_i\phi_i^S({\pmb r}_i) \ .
\end{equation}

The last term of Eq.~\ref{elec_pot_E} is added in order to remove the prime over the summation in the third term of Eq.~\ref{elec_pot_E} and corresponds to the potential produced by the i'th Gaussian charge,
\begin{eqnarray}
\phi_i^{self}({\pmb r})=q_i \dfrac{\text{erf}(\kappa_e|{\pmb r}-{\pmb r_i}|)}{\epsilon_w|{\pmb r}-{\pmb r_i}|} \ .
\end{eqnarray}
The total self energy is
\begin{equation}
U_{self}=-\frac{1}{2}\sum_{i=1}^{N}q_i\phi_i^{self}({\pmb r}_i)=-\dfrac{\kappa_e}{\epsilon_w\sqrt{\pi}}\sum_{i=1}^Nq_i^2 \ .
\end{equation}

The third and fourth terms of Eq.~\ref{elec_pot_E} define the long range potential, $\phi_i^L({\pmb r})$,
\begin{eqnarray}\label{philr}
\phi^L({\pmb r})=\sum_{\pmb n}^{\infty}\sum_{j=1}^{N}\text{q}_j\frac{\text{erf}(\kappa_e|{\pmb r}-{\pmb r}_j-{\pmb r}_{rep}|)}{\epsilon_w |{\pmb r}-{\pmb r}_j-{\pmb r}_{rep}|} + \nonumber \\
\sum_{\pmb n}^{\infty}\sum_{j=1}^{N}\gamma\text{q}_j\frac{\text{erf}(\kappa_e|{\pmb r}-{\pmb r}_j'-{\pmb r}_{rep}|)}{\epsilon_w |{\pmb r}-{\pmb r}_j'-{\pmb r}_{rep}|} \ .
\end{eqnarray}
We can Fourier transform Eq. \ref{philr}, resulting in
\begin{eqnarray}\label{elec_pot_E_f}
\phi^L({\pmb r})=\sum_{{\pmb k}={\pmb 0}}^{\infty}\sum_{j=1}^{N}\frac{4\pi \text{q}_j}{\epsilon_w V |{\pmb k}|^2}\exp{[-\frac{|{\pmb k}|^2}{4\kappa_e^2}]}\Bigg[\exp{[i{\pmb k}\cdot({\pmb r}-{\pmb r}_j)]} + \nonumber \\
\gamma\exp{[i{\pmb k}\cdot({\pmb r}-{\pmb r}_j')]} \Bigg] \ ,
\end{eqnarray}
with ${\pmb k}=(\frac{2\pi}{L_x}n_x,\frac{2\pi}{L_y}n_y,\frac{2\pi}{L_z}n_z)$. We note  that the term corresponding to ${\pmb k}=(0,0,0)$ is divergent. However, the divergence can be renormalized away by changing the zero point of the potential, as discussed in Ref.\cite{DoGi16}. We expand the singular term around ${\pmb k}=(0,0,0)$ and keep the non-vanishing factors
\begin{eqnarray}\label{pot}
\lim_{{\pmb k} \rightarrow 0}\sum_{j=1}^{N}\frac{4\pi \text{q}_j}{\epsilon_w V}\Bigg[\frac{1}{|{\pmb k}|^2}-\frac{1}{4\kappa_e^2}+ \nonumber \\
\frac{i{\pmb k}\cdot({\pmb r}-{\pmb r}_j)}{|{\pmb k}|^2} + \gamma\frac{i{\pmb k}\cdot({\pmb r}-{\pmb r}_j')}{|{\pmb k}|^2} - \nonumber \\
\dfrac{[{\pmb k}\cdot({\pmb r}-{\pmb r}_j)]^2}{2|{\pmb k}|^2} - \gamma\dfrac{[{\pmb k}\cdot({\pmb r}-{\pmb r}_j')]^2}{2|{\pmb k}|^2}\Bigg]  \ .
\end{eqnarray}
The first two terms are zero for neutral systems, $\sum_{j=1}^N q_j=0$, but diverge for systems with net charge. However, they are independent of position and can be renormalized away by simply redefining the zero of the potential~\cite{DoGi16}. The second and third terms are zero, as shown in Ref.~\cite{DoGi16}. The remaining terms can be calculated taking into account the aspect ratio of the infinite system. For details of calculations see Ref.~\cite{DoGi16}. For a slab geometry, the directions   $\hat{\pmb x}$ and $\hat{\pmb y}$ go to infinity much faster than  $\hat{\pmb z}$,  resulting in finite correction potential 
\begin{equation}
\phi_i^{cor}({\pmb r})= -\sum_{j=1}^N\frac{2\pi q_j}{\epsilon_w V}\Bigg[(z-z_j)^2+\gamma(z-z_j')^2\Bigg] \ .
\end{equation}
The correction energy is $U_{cor}=(1/2)\sum_{i=1}^{N}q_i\phi_i^{cor}({\pmb r_i})$, which after a short calculation reduces to
\begin{equation}
U_{cor}=\frac{2\pi}{\epsilon_w V} \left[ M_z^2 (1-\gamma) - G_z Q_t (1+\gamma) \right] \ ,
\end{equation}
where
\begin{equation}
M_z=\sum_{i=1}^{N}q_i z_i \ \ \text{,} \ \ G_z=\sum_{i=1}^{N}q_i z_i^2 \ \ \text{and} \ \ Q_t=\sum_{i=1}^{N}q_i \ .
\end{equation}

We can now exclude ${\pmb k}={\pmb 0}$ in the long range potential, Eq.~$\ref{elec_pot_E_f}$, since it is now accounted for by the correction potential. The long range energy is given by $U_{L}=(1/2)\sum_{i=1}^{N}q_i\phi_i^L({\pmb r_i})$ which can be written as
\begin{eqnarray}\label{U_long}
U_L = \sum_{{\pmb k}\neq{\pmb 0}}\frac{2\pi}{\epsilon_w V |{\pmb k}|^2}
\text{exp}(-\frac{|{\pmb k}|^2}{4\kappa_e^2}) \times \nonumber \\
\left[A({\pmb k})^2 + B({\pmb k})^2 + A({\pmb k}) C({\pmb k}) + B({\pmb k}) D({\pmb k})\right] \ ,
\end{eqnarray}
where
\begin{eqnarray}
A({\pmb k})= \sum_{i=1}^{N}q_i \cos{({\pmb k}\cdot {\pmb r}_i)} \ , \nonumber \\
B({\pmb k})=- \sum_{i=1}^{N}q_i \sin{({\pmb k}\cdot {\pmb r}_i)} \ , \nonumber \\
C({\pmb k})= \sum_{i=1}^{N}\gamma q_i \cos{({\pmb k}\cdot {\pmb r}'_i)} \ , \nonumber \\
D({\pmb k})=- \sum_{i=1}^{N} \gamma q_i \sin{({\pmb k}\cdot {\pmb r}'_i)} \nonumber \ .
\end{eqnarray}

In general, charged surfaces also contribute with counterions, which must be included in the main simulation cell. The electrostatic potential produced by the surface acts on all the charged 
particles inside the  simulation cell and can be added as an external field,
\begin{eqnarray}
\phi_P({\pmb r}) =-\frac{2\pi}{\epsilon_w}(1+\gamma)\sigma z \ ,
\end{eqnarray}
where $\sigma$ is the surface charge density. The total plate-ions interaction energy is $U_P=\sum_{i=1}^N q_i \phi_P({\pmb r}_i)$.

The total electrostatic energy is given by sum of all contributions,
\begin{equation}
U=U_S+U_{self}+U_{cor}+U_L+U_P \ .
\end{equation}

\bibliography{ref}

\providecommand*\mcitethebibliography{\thebibliography}
\csname @ifundefined\endcsname{endmcitethebibliography}
  {\let\endmcitethebibliography\endthebibliography}{}
\begin{mcitethebibliography}{64}
\providecommand*\natexlab[1]{#1}
\providecommand*\mciteSetBstSublistMode[1]{}
\providecommand*\mciteSetBstMaxWidthForm[2]{}
\providecommand*\mciteBstWouldAddEndPuncttrue
  {\def\EndOfBibitem{\unskip.}}
\providecommand*\mciteBstWouldAddEndPunctfalse
  {\let\EndOfBibitem\relax}
\providecommand*\mciteSetBstMidEndSepPunct[3]{}
\providecommand*\mciteSetBstSublistLabelBeginEnd[3]{}
\providecommand*\EndOfBibitem{}
\mciteSetBstSublistMode{f}
\mciteSetBstMaxWidthForm{subitem}{(\alph{mcitesubitemcount})}
\mciteSetBstSublistLabelBeginEnd
  {\mcitemaxwidthsubitemform\space}
  {\relax}
  {\relax}

\bibitem[Bajdur et~al.(2002)Bajdur, Pajaczkowska, Makarucha, Sulkowska, and
  Sulkowski]{BaPa02}
Bajdur,~W.; Pajaczkowska,~J.; Makarucha,~B.; Sulkowska,~A.; Sulkowski,~W.~W.
  Effective Polyelectrolytes Synthesised from Expanded Polystyrene Wastes.
  \emph{Eur. Polym. J.} \textbf{2002}, \emph{38}, 299--304\relax
\mciteBstWouldAddEndPuncttrue
\mciteSetBstMidEndSepPunct{\mcitedefaultmidpunct}
{\mcitedefaultendpunct}{\mcitedefaultseppunct}\relax
\EndOfBibitem
\bibitem[Wang et~al.(2013)Wang, Janout, and Regen]{WaJa13}
Wang,~M.~H.; Janout,~V.; Regen,~S.~L. Gas Transport across Hyperthin Membranes.
  \emph{Acc. Chem. Res.} \textbf{2013}, \emph{46}, 2743--2754\relax
\mciteBstWouldAddEndPuncttrue
\mciteSetBstMidEndSepPunct{\mcitedefaultmidpunct}
{\mcitedefaultendpunct}{\mcitedefaultseppunct}\relax
\EndOfBibitem
\bibitem[Cakmak et~al.(2004)Cakmak, Ulukanli, Tuzcu, Karabuga, and
  Genctav]{CaUl04}
Cakmak,~I.; Ulukanli,~Z.; Tuzcu,~M.; Karabuga,~S.; Genctav,~K. Synthesis and
  Characterization of Novel Antimicrobial Cationic Polyelectrolytes. \emph{Eur.
  Polym. J.} \textbf{2004}, \emph{40}, 2373--2379\relax
\mciteBstWouldAddEndPuncttrue
\mciteSetBstMidEndSepPunct{\mcitedefaultmidpunct}
{\mcitedefaultendpunct}{\mcitedefaultseppunct}\relax
\EndOfBibitem
\bibitem[Lu et~al.(2005)Lu, Rininsland, Wittenburg, Achyuthan, McBranch, and
  Whitten]{LuRi05}
Lu,~L.~D.; Rininsland,~F.~H.; Wittenburg,~S.~K.; Achyuthan,~K.~E.;
  McBranch,~D.~W.; Whitten,~D.~G. Biocidal Activity of a Light-Absorbing
  Fluorescent Conjugated Polyelectrolyte. \emph{Langmuir} \textbf{2005},
  \emph{21}, 10154--10159\relax
\mciteBstWouldAddEndPuncttrue
\mciteSetBstMidEndSepPunct{\mcitedefaultmidpunct}
{\mcitedefaultendpunct}{\mcitedefaultseppunct}\relax
\EndOfBibitem
\bibitem[Kiss et~al.(2012)Kiss, Heine, Hill, He, Keusgen, P\'enzes,
  Schn\"oller, Gyulai, Mendrek, Keul, and et~al]{KiHe12}
Kiss,~E.; Heine,~E.~T.; Hill,~K.; He,~Y.~C.; Keusgen,~N.; P\'enzes,~C.~B.;
  Schn\"oller,~D.; Gyulai,~G.; Mendrek,~A.; Keul,~H.; et~al, Membrane Affinity
  and Antibacterial Properties of Cationic Polyelectrolytes with Different
  Hydrophobicity. \emph{Macromol. Biosci.} \textbf{2012}, \emph{12},
  1181--1189\relax
\mciteBstWouldAddEndPuncttrue
\mciteSetBstMidEndSepPunct{\mcitedefaultmidpunct}
{\mcitedefaultendpunct}{\mcitedefaultseppunct}\relax
\EndOfBibitem
\bibitem[Hoque et~al.(2012)Hoque, Akkapeddi, Yarlagadda, Uppu, Kumar, and
  Haldar]{HoAk12}
Hoque,~J.; Akkapeddi,~P.; Yarlagadda,~V.; Uppu,~D. S. S.~M.; Kumar,~P.;
  Haldar,~J. Cleavable Cationic Antibacterial Amphiphiles: Synthesis, Mechanism
  of Action, and Cytotoxicities. \emph{Langmuir} \textbf{2012}, \emph{28},
  12225--12234\relax
\mciteBstWouldAddEndPuncttrue
\mciteSetBstMidEndSepPunct{\mcitedefaultmidpunct}
{\mcitedefaultendpunct}{\mcitedefaultseppunct}\relax
\EndOfBibitem
\bibitem[Falentin-Daudr\'e et~al.(2012)Falentin-Daudr\'e, Faure, Svaldo-Lanero,
  Farina, J\'er\^ome, {Van De Weerdt}, Martial, Duwez, and Detrembleur]{FaFa12}
Falentin-Daudr\'e,~C.; Faure,~E.; Svaldo-Lanero,~T.; Farina,~F.;
  J\'er\^ome,~C.; {Van De Weerdt},~C.; Martial,~J.; Duwez,~A.~S.;
  Detrembleur,~C. Antibacterial Polyelectrolyte Micelles for Coating Stainless
  Steel. \emph{Langmuir} \textbf{2012}, \emph{28}, 7233--7241\relax
\mciteBstWouldAddEndPuncttrue
\mciteSetBstMidEndSepPunct{\mcitedefaultmidpunct}
{\mcitedefaultendpunct}{\mcitedefaultseppunct}\relax
\EndOfBibitem
\bibitem[Feng et~al.(2010)Feng, Liu, Wang, and Zhu]{FeLi10}
Feng,~X.~L.; Liu,~L.~B.; Wang,~S.; Zhu,~D.~B. Water-Soluble Fluorescent
  Conjugated Polymers and their Interactions with Biomacromolecules for
  Sensitive Sensors. \emph{Chem. Soc. Rev.} \textbf{2010}, \emph{39},
  2411--2419\relax
\mciteBstWouldAddEndPuncttrue
\mciteSetBstMidEndSepPunct{\mcitedefaultmidpunct}
{\mcitedefaultendpunct}{\mcitedefaultseppunct}\relax
\EndOfBibitem
\bibitem[Kim et~al.(2011)Kim, Guo, Zhu, Yoon, and Tian]{KiGu11}
Kim,~H.~N.; Guo,~Z.~Q.; Zhu,~W.~H.; Yoon,~J.; Tian,~H. Recent Progress on
  Polymer-Based Fluorescent and Colorimetric Chemosensors. \emph{Chem. Soc.
  Rev.} \textbf{2011}, \emph{40}, 79--93\relax
\mciteBstWouldAddEndPuncttrue
\mciteSetBstMidEndSepPunct{\mcitedefaultmidpunct}
{\mcitedefaultendpunct}{\mcitedefaultseppunct}\relax
\EndOfBibitem
\bibitem[Kong and Muthukumar(1998)Kong, and Muthukumar]{KoMu98}
Kong,~C.~Y.; Muthukumar,~M. Monte Carlo Study of a Polyelectrolyte onto Charged
  Surfaces. \emph{J. Chem. Phys.} \textbf{1998}, \emph{109}, 1522--1557\relax
\mciteBstWouldAddEndPuncttrue
\mciteSetBstMidEndSepPunct{\mcitedefaultmidpunct}
{\mcitedefaultendpunct}{\mcitedefaultseppunct}\relax
\EndOfBibitem
\bibitem[Chodanowski and Stoll(2001)Chodanowski, and Stoll]{ChSt01}
Chodanowski,~P.; Stoll,~S. Polyelectrolyte Adsorption on Charged Particles:
  Ionic Concentration and Particle Size Effects - A Monte Carlo Approach.
  \emph{J. Chem. Phys.} \textbf{2001}, \emph{115}, 4951--4960\relax
\mciteBstWouldAddEndPuncttrue
\mciteSetBstMidEndSepPunct{\mcitedefaultmidpunct}
{\mcitedefaultendpunct}{\mcitedefaultseppunct}\relax
\EndOfBibitem
\bibitem[Dobrynin et~al.(2000)Dobrynin, Deshkovski, and Rubinstein]{DoDe00}
Dobrynin,~A.~V.; Deshkovski,~A.; Rubinstein,~M. Adsorption of Polyelectrolytes
  at an Oppositely Charged Surface. \emph{Phys. Rev. Lett.} \textbf{2000},
  \emph{84}, 3101--3104\relax
\mciteBstWouldAddEndPuncttrue
\mciteSetBstMidEndSepPunct{\mcitedefaultmidpunct}
{\mcitedefaultendpunct}{\mcitedefaultseppunct}\relax
\EndOfBibitem
\bibitem[Dobrynin et~al.(2001)Dobrynin, Deshkovski, and Rubinstein]{DoDe01}
Dobrynin,~A.~V.; Deshkovski,~A.; Rubinstein,~M. Adsorption of Polyelectrolytes
  at Oppositely Charged Surfaces. \emph{Macromolecules} \textbf{2001},
  \emph{34}, 3421--3436\relax
\mciteBstWouldAddEndPuncttrue
\mciteSetBstMidEndSepPunct{\mcitedefaultmidpunct}
{\mcitedefaultendpunct}{\mcitedefaultseppunct}\relax
\EndOfBibitem
\bibitem[Dobrynin and Rubinstein(2002)Dobrynin, and Rubinstein]{DoRu02}
Dobrynin,~A.~V.; Rubinstein,~M. Adsorption of Hydrophobic Polyelectrolytes at
  Oppositely Charged Surfaces. \emph{Macromolecules} \textbf{2002}, \emph{35},
  2754--2768\relax
\mciteBstWouldAddEndPuncttrue
\mciteSetBstMidEndSepPunct{\mcitedefaultmidpunct}
{\mcitedefaultendpunct}{\mcitedefaultseppunct}\relax
\EndOfBibitem
\bibitem[McNamara et~al.(2002)McNamara, Kong, and Muthukumar]{McKo02}
McNamara,~J.; Kong,~C.~Y.; Muthukumar,~M. Monte Carlo Studies of Adsorption of
  a Sequenced Polyelectrolyte to Patterned Surfaces. \emph{J. Chem. Phys.}
  \textbf{2002}, \emph{117}, 5354--5360\relax
\mciteBstWouldAddEndPuncttrue
\mciteSetBstMidEndSepPunct{\mcitedefaultmidpunct}
{\mcitedefaultendpunct}{\mcitedefaultseppunct}\relax
\EndOfBibitem
\bibitem[Reddy and Yethiraj(2010)Reddy, and Yethiraj]{ReYe10}
Reddy,~G.; Yethiraj,~A. Solvent Effects in Polyelectrolyte Adsorption: Computer
  Simulations with Explicit and Implicit Solvent. \emph{J. Chem. Phys.}
  \textbf{2010}, \emph{132}, 074903\relax
\mciteBstWouldAddEndPuncttrue
\mciteSetBstMidEndSepPunct{\mcitedefaultmidpunct}
{\mcitedefaultendpunct}{\mcitedefaultseppunct}\relax
\EndOfBibitem
\bibitem[Qiao et~al.(2011)Qiao, Cerd\`a, and Holm]{QiCe11}
Qiao,~B.; Cerd\`a,~J.~J.; Holm,~C. Atomistic Study of Surface Effects on
  Polyelectrolyte Adsorption: Case Study of a Poly(styrenesulfonate) Monolayer.
  \emph{Macromolecules} \textbf{2011}, \emph{44}, 1707–1718\relax
\mciteBstWouldAddEndPuncttrue
\mciteSetBstMidEndSepPunct{\mcitedefaultmidpunct}
{\mcitedefaultendpunct}{\mcitedefaultseppunct}\relax
\EndOfBibitem
\bibitem[Faraudo and {Martin-Molina}(2013)Faraudo, and {Martin-Molina}]{FaMa13}
Faraudo,~J.; {Martin-Molina},~A. Competing Forces in the Interaction of
  Polyelectrolytes with Charged Surfaces. \emph{Curr. Opin. Colloid Interface
  Sci.} \textbf{2013}, \emph{18}, 517--523\relax
\mciteBstWouldAddEndPuncttrue
\mciteSetBstMidEndSepPunct{\mcitedefaultmidpunct}
{\mcitedefaultendpunct}{\mcitedefaultseppunct}\relax
\EndOfBibitem
\bibitem[Stoll and Chodanowski(2002)Stoll, and Chodanowski]{StCh02}
Stoll,~S.; Chodanowski,~P. Polyelectrolyte Adsorption on an Oppositely Charged
  Spherical Particle. Chain Rigidity Effects. \emph{Macromolecules}
  \textbf{2002}, \emph{35}, 9556–9562\relax
\mciteBstWouldAddEndPuncttrue
\mciteSetBstMidEndSepPunct{\mcitedefaultmidpunct}
{\mcitedefaultendpunct}{\mcitedefaultseppunct}\relax
\EndOfBibitem
\bibitem[Messina et~al.(2004)Messina, Holm, and Kremer]{MeHo04}
Messina,~R.; Holm,~C.; Kremer,~K. Polyelectrolyte Adsorption and Multilayering
  on Charged Colloidal Particles. \emph{J. Polym. Sci. Pol. Phys.}
  \textbf{2004}, \emph{42}, 3557--3570\relax
\mciteBstWouldAddEndPuncttrue
\mciteSetBstMidEndSepPunct{\mcitedefaultmidpunct}
{\mcitedefaultendpunct}{\mcitedefaultseppunct}\relax
\EndOfBibitem
\bibitem[Salomaki and Kankare(2008)Salomaki, and Kankare]{SaKa08}
Salomaki,~M.; Kankare,~J. Specific Anion Effect in Swelling of Polyelectrolyte
  Multilayers. \emph{Macromolecules} \textbf{2008}, \emph{41}, 4423--4428\relax
\mciteBstWouldAddEndPuncttrue
\mciteSetBstMidEndSepPunct{\mcitedefaultmidpunct}
{\mcitedefaultendpunct}{\mcitedefaultseppunct}\relax
\EndOfBibitem
\bibitem[{dos Santos} and Levin(2013){dos Santos}, and Levin]{DoLe13}
{dos Santos},~A.~P.; Levin,~Y. Adsorption of Cationic Polyions onto a
  Hydrophobic Surface in the Presence of Hofmeister Salts. \emph{Soft Matter}
  \textbf{2013}, \emph{9}, 10545--10549\relax
\mciteBstWouldAddEndPuncttrue
\mciteSetBstMidEndSepPunct{\mcitedefaultmidpunct}
{\mcitedefaultendpunct}{\mcitedefaultseppunct}\relax
\EndOfBibitem
\bibitem[Kou et~al.(2015)Kou, Zhang, Wang, and Liu]{KoZh15}
Kou,~R.; Zhang,~J.; Wang,~T.; Liu,~G. Interactions between Polyelectrolyte
  Brushes and Hofmeister Ions: Chaotropes versus Kosmotropes. \emph{Langmuir}
  \textbf{2015}, \emph{31}, 10461--10468\relax
\mciteBstWouldAddEndPuncttrue
\mciteSetBstMidEndSepPunct{\mcitedefaultmidpunct}
{\mcitedefaultendpunct}{\mcitedefaultseppunct}\relax
\EndOfBibitem
\bibitem[Ha and Liu(1999)Ha, and Liu]{HaLi99}
Ha,~B.~Y.; Liu,~A.~J. Counterion-Mediated, Non-Pairwise-Additive Attractions in
  Bundles of Like-Charged Rods. \emph{Phys. Rev. E} \textbf{1999}, \emph{60},
  803--813\relax
\mciteBstWouldAddEndPuncttrue
\mciteSetBstMidEndSepPunct{\mcitedefaultmidpunct}
{\mcitedefaultendpunct}{\mcitedefaultseppunct}\relax
\EndOfBibitem
\bibitem[Arenzon et~al.(1999)Arenzon, Stilck, and Levin]{ArSt99}
Arenzon,~J.~J.; Stilck,~J.~F.; Levin,~Y. Simple Model for Attraction between
  Like-Charged Polyions. \emph{Eur. Phys. J. B} \textbf{1999}, \emph{12},
  79--82\relax
\mciteBstWouldAddEndPuncttrue
\mciteSetBstMidEndSepPunct{\mcitedefaultmidpunct}
{\mcitedefaultendpunct}{\mcitedefaultseppunct}\relax
\EndOfBibitem
\bibitem[Deserno et~al.(2001)Deserno, {Jim\'enez-\'Angeles}, Holm, and
  {Lozada-Cassou}]{DeJi01}
Deserno,~M.; {Jim\'enez-\'Angeles},~F.; Holm,~C.; {Lozada-Cassou},~M.
  Overcharging of DNA in the Presence of Salt: Theory and Simulation. \emph{J.
  Phys. Chem. B} \textbf{2001}, \emph{105}, 10983--10991\relax
\mciteBstWouldAddEndPuncttrue
\mciteSetBstMidEndSepPunct{\mcitedefaultmidpunct}
{\mcitedefaultendpunct}{\mcitedefaultseppunct}\relax
\EndOfBibitem
\bibitem[Butler et~al.(2003)Butler, Angelini, Tang, and Wong]{BuAn03}
Butler,~J.~C.; Angelini,~T.; Tang,~J.~X.; Wong,~G. C.~L. Ion Multivalence and
  Like-Charged Polyelectrolyte Attraction. \emph{Phys. Rev. Lett.}
  \textbf{2003}, \emph{91}, 028301--028303\relax
\mciteBstWouldAddEndPuncttrue
\mciteSetBstMidEndSepPunct{\mcitedefaultmidpunct}
{\mcitedefaultendpunct}{\mcitedefaultseppunct}\relax
\EndOfBibitem
\bibitem[Deserno et~al.(2003)Deserno, Arnold, and Holm]{DeAr03}
Deserno,~M.; Arnold,~A.; Holm,~C. Attraction and Ionic Correlations between
  Charged Stiff Polyelectrolytes. \emph{Macromolecules} \textbf{2003},
  \emph{36}, 249--259\relax
\mciteBstWouldAddEndPuncttrue
\mciteSetBstMidEndSepPunct{\mcitedefaultmidpunct}
{\mcitedefaultendpunct}{\mcitedefaultseppunct}\relax
\EndOfBibitem
\bibitem[Angelini et~al.(2003)Angelini, Liang, Wriggers, and Wong]{AnLi03}
Angelini,~T.~E.; Liang,~H.; Wriggers,~W.; Wong,~G. C.~L. Like-Charge Attraction
  between Polyelectrolytes Induced by Counterion Charge Density Waves.
  \emph{Proc. Natl. Acad. Sci. U. S. A.} \textbf{2003}, \emph{100},
  8634--8637\relax
\mciteBstWouldAddEndPuncttrue
\mciteSetBstMidEndSepPunct{\mcitedefaultmidpunct}
{\mcitedefaultendpunct}{\mcitedefaultseppunct}\relax
\EndOfBibitem
\bibitem[Molnar and Rieger(2005)Molnar, and Rieger]{MoRi05}
Molnar,~F.; Rieger,~J. ``Like-Charge Attraction'' between Anionic
  Polyelectrolytes: Molecular Dynamics Simulations. \emph{Langmuir}
  \textbf{2005}, \emph{21}, 786--789\relax
\mciteBstWouldAddEndPuncttrue
\mciteSetBstMidEndSepPunct{\mcitedefaultmidpunct}
{\mcitedefaultendpunct}{\mcitedefaultseppunct}\relax
\EndOfBibitem
\bibitem[Turesson et~al.(2011)Turesson, Labbez, and Nonat]{TuLa11}
Turesson,~M.; Labbez,~C.; Nonat,~A. Calcium Mediated Polyelectrolyte Adsorption
  on Like-Charged Surfaces. \emph{Langmuir} \textbf{2011}, \emph{27},
  13572--13581\relax
\mciteBstWouldAddEndPuncttrue
\mciteSetBstMidEndSepPunct{\mcitedefaultmidpunct}
{\mcitedefaultendpunct}{\mcitedefaultseppunct}\relax
\EndOfBibitem
\bibitem[{Luque-Caballero} et~al.(2014){Luque-Caballero}, {Mart\'in-Molina},
  and {Quesada-P\'erez}]{LuMa14}
{Luque-Caballero},~G.; {Mart\'in-Molina},~A.; {Quesada-P\'erez},~M.
  Polyelectrolyte Adsorption onto Like-Charged Surfaces Mediated by Trivalent
  Counterions: A Monte Carlo Simulation Study. \emph{J. Chem. Phys.}
  \textbf{2014}, \emph{140}, 174701--174710\relax
\mciteBstWouldAddEndPuncttrue
\mciteSetBstMidEndSepPunct{\mcitedefaultmidpunct}
{\mcitedefaultendpunct}{\mcitedefaultseppunct}\relax
\EndOfBibitem
\bibitem[Tiraferri et~al.(2015)Tiraferri, Maroni, and Borkovec]{TiMa15}
Tiraferri,~A.; Maroni,~P.; Borkovec,~M. Adsorption of Polyelectrolyte to
  Like-Charged Substrates Induced by Multivalent Counterions as Exemplified by
  Poly(styrene sulfonate) and Silica. \emph{Phys. Chem. Chem. Phys.}
  \textbf{2015}, \emph{17}, 10348--10352\relax
\mciteBstWouldAddEndPuncttrue
\mciteSetBstMidEndSepPunct{\mcitedefaultmidpunct}
{\mcitedefaultendpunct}{\mcitedefaultseppunct}\relax
\EndOfBibitem
\bibitem[{Allen, M. P. and Tildesley, D. J.}(1987)]{AlTi87}
{Allen, M. P. and Tildesley, D. J.}, \emph{Computer Simulations of Liquids};
  Oxford: Oxford University Press: New York, 1987\relax
\mciteBstWouldAddEndPuncttrue
\mciteSetBstMidEndSepPunct{\mcitedefaultmidpunct}
{\mcitedefaultendpunct}{\mcitedefaultseppunct}\relax
\EndOfBibitem
\bibitem[Torrie and Valleau(1980)Torrie, and Valleau]{ToVa80}
Torrie,~G.~M.; Valleau,~J.~P. Electrical Double Layers. I. Monte Carlo Study of
  a Uniformly Charged Surface. \emph{J. Chem. Phys.} \textbf{1980}, \emph{73},
  5807--5816\relax
\mciteBstWouldAddEndPuncttrue
\mciteSetBstMidEndSepPunct{\mcitedefaultmidpunct}
{\mcitedefaultendpunct}{\mcitedefaultseppunct}\relax
\EndOfBibitem
\bibitem[Torrie et~al.(1984)Torrie, Valleau, and Outhwaite]{ToVa84}
Torrie,~G.~M.; Valleau,~J.~P.; Outhwaite,~C.~W. Electrical Double Layers. VI.
  Image effects for Divalent Ions. \emph{J. Chem. Phys.} \textbf{1984},
  \emph{81}, 6296--6300\relax
\mciteBstWouldAddEndPuncttrue
\mciteSetBstMidEndSepPunct{\mcitedefaultmidpunct}
{\mcitedefaultendpunct}{\mcitedefaultseppunct}\relax
\EndOfBibitem
\bibitem[Lekner(1991)]{Le91}
Lekner,~J. Summation of Coulomb Fields in Computer-Simulated Disordered
  Systems. \emph{Physica A} \textbf{1991}, \emph{176}, 485--498\relax
\mciteBstWouldAddEndPuncttrue
\mciteSetBstMidEndSepPunct{\mcitedefaultmidpunct}
{\mcitedefaultendpunct}{\mcitedefaultseppunct}\relax
\EndOfBibitem
\bibitem[Widmann and Adolf(1997)Widmann, and Adolf]{WiAd97}
Widmann,~A.~H.; Adolf,~D.~B. A Comparison of Ewald Summation Techniques for
  Planar Surfaces. \emph{Comput. Phys. Commun.} \textbf{1997}, \emph{107},
  167--186\relax
\mciteBstWouldAddEndPuncttrue
\mciteSetBstMidEndSepPunct{\mcitedefaultmidpunct}
{\mcitedefaultendpunct}{\mcitedefaultseppunct}\relax
\EndOfBibitem
\bibitem[Yeh and Berkowitz(1999)Yeh, and Berkowitz]{YeBe99}
Yeh,~I.; Berkowitz,~M.~L. Ewald Summation for Systems with Slab Geometry.
  \emph{J. Chem. Phys.} \textbf{1999}, \emph{111}, 3155--3162\relax
\mciteBstWouldAddEndPuncttrue
\mciteSetBstMidEndSepPunct{\mcitedefaultmidpunct}
{\mcitedefaultendpunct}{\mcitedefaultseppunct}\relax
\EndOfBibitem
\bibitem[Arnold and Holm(2002)Arnold, and Holm]{ArHo02}
Arnold,~A.; Holm,~C. A Novel Method for Calculating Electrostatic Interaction
  in 2D Periodic Slab Geometry. \emph{Chem. Phys. Lett.} \textbf{2002},
  \emph{354}, 324--330\relax
\mciteBstWouldAddEndPuncttrue
\mciteSetBstMidEndSepPunct{\mcitedefaultmidpunct}
{\mcitedefaultendpunct}{\mcitedefaultseppunct}\relax
\EndOfBibitem
\bibitem[{dos Santos} et~al.(2016){dos Santos}, Girotto, and Levin]{DoGi16}
{dos Santos},~A.~P.; Girotto,~M.; Levin,~Y. Simulations of Coulomb Systems with
  Slab Geometry Using an Efficient 3D Ewald Summation Method. \emph{J. Chem.
  Phys.} \textbf{2016}, \emph{144}, 144103--144109\relax
\mciteBstWouldAddEndPuncttrue
\mciteSetBstMidEndSepPunct{\mcitedefaultmidpunct}
{\mcitedefaultendpunct}{\mcitedefaultseppunct}\relax
\EndOfBibitem
\bibitem[Ballenegger et~al.(2009)Ballenegger, Arnold, and Cerd\`a]{BaAr09}
Ballenegger,~V.; Arnold,~A.; Cerd\`a,~J.~J. Simulation of Non-Neutral Slab
  Systems with Long-Range Electrostatic Interactions in Two-Dimensional
  Periodic Boundary Conditions. \emph{J. Chem. Phys.} \textbf{2009},
  \emph{131}, 094107--094116\relax
\mciteBstWouldAddEndPuncttrue
\mciteSetBstMidEndSepPunct{\mcitedefaultmidpunct}
{\mcitedefaultendpunct}{\mcitedefaultseppunct}\relax
\EndOfBibitem
\bibitem[Ballenegger(2014)]{Ba14}
Ballenegger,~V. Communication: On the Origin of the Surface Term in the Ewald
  Formula. \emph{J. Chem. Phys.} \textbf{2014}, \emph{140},
  161102--161105\relax
\mciteBstWouldAddEndPuncttrue
\mciteSetBstMidEndSepPunct{\mcitedefaultmidpunct}
{\mcitedefaultendpunct}{\mcitedefaultseppunct}\relax
\EndOfBibitem
\bibitem[{{dos Santos}, A. P. and Levin, Y.}(2014)]{DoLe14b}
{{dos Santos}, A. P. and Levin, Y.}, \emph{Electrostatics of Soft and
  Disordered Matter}; CRC Press: Boca Raton, 2014; Chapter 14\relax
\mciteBstWouldAddEndPuncttrue
\mciteSetBstMidEndSepPunct{\mcitedefaultmidpunct}
{\mcitedefaultendpunct}{\mcitedefaultseppunct}\relax
\EndOfBibitem
\bibitem[Dias et~al.(2005)Dias, Pais, Linse, Miguel, and Lindman]{DiPa05}
Dias,~R.~S.; Pais,~A. A. C.~C.; Linse,~P.; Miguel,~M.~G.; Lindman,~B. Polyion
  Adsorption onto Catanionic Surfaces. A Monte Carlo Study. \emph{J. Phys.
  Chem. B} \textbf{2005}, \emph{109}, 11781--11788\relax
\mciteBstWouldAddEndPuncttrue
\mciteSetBstMidEndSepPunct{\mcitedefaultmidpunct}
{\mcitedefaultendpunct}{\mcitedefaultseppunct}\relax
\EndOfBibitem
\bibitem[{Quesada-P\'erez} and {Mart\'in-Molina}(2013){Quesada-P\'erez}, and
  {Mart\'in-Molina}]{QuMa13}
{Quesada-P\'erez},~M.; {Mart\'in-Molina},~A. Monte Carlo Simulation of
  Thermo-Responsive Charged Nanogels in Salt-Free Solutions. \emph{Soft Matter}
  \textbf{2013}, \emph{9}, 7086--7094\relax
\mciteBstWouldAddEndPuncttrue
\mciteSetBstMidEndSepPunct{\mcitedefaultmidpunct}
{\mcitedefaultendpunct}{\mcitedefaultseppunct}\relax
\EndOfBibitem
\bibitem[Kanduc et~al.(2010)Kanduc, Naji, and Podgornik]{KaNa10}
Kanduc,~M.; Naji,~A.; Podgornik,~R. Counterion-Mediated Weak and Strong
  Coupling Electrostatic Interaction between Like-Charged Cylindrical
  Dielectrics. \emph{J. Chem. Phys.} \textbf{2010}, \emph{132},
  224703--224720\relax
\mciteBstWouldAddEndPuncttrue
\mciteSetBstMidEndSepPunct{\mcitedefaultmidpunct}
{\mcitedefaultendpunct}{\mcitedefaultseppunct}\relax
\EndOfBibitem
\bibitem[Nagy et~al.(2011)Nagy, Henderson, and Boda]{NaHe11}
Nagy,~T.; Henderson,~D.; Boda,~D. Simulation of an Electrical Double Layer
  Model with a Low Dielectric Layer between the Electrode and the Electrolyte.
  \emph{J. Phys. Chem. B} \textbf{2011}, \emph{115}, 11409--11419\relax
\mciteBstWouldAddEndPuncttrue
\mciteSetBstMidEndSepPunct{\mcitedefaultmidpunct}
{\mcitedefaultendpunct}{\mcitedefaultseppunct}\relax
\EndOfBibitem
\bibitem[Samaj and Trizac(2011)Samaj, and Trizac]{SaTr11}
Samaj,~L.; Trizac,~E. Counterions at Highly Charged Interfaces: From One Plate
  to Like-Charge Attraction. \emph{Phys. Rev. Lett.} \textbf{2011}, \emph{106},
  078301--078304\relax
\mciteBstWouldAddEndPuncttrue
\mciteSetBstMidEndSepPunct{\mcitedefaultmidpunct}
{\mcitedefaultendpunct}{\mcitedefaultseppunct}\relax
\EndOfBibitem
\bibitem[Bakhshandeh et~al.(2011)Bakhshandeh, {dos Santos}, and Levin]{BaDo11}
Bakhshandeh,~A.; {dos Santos},~A.~P.; Levin,~Y. Weak and Strong Coupling
  Theories for Polarizable Colloids and Nanoparticles. \emph{Phys. Rev. Lett.}
  \textbf{2011}, \emph{107}, 107801--107805\relax
\mciteBstWouldAddEndPuncttrue
\mciteSetBstMidEndSepPunct{\mcitedefaultmidpunct}
{\mcitedefaultendpunct}{\mcitedefaultseppunct}\relax
\EndOfBibitem
\bibitem[{dos Santos} et~al.(2011){dos Santos}, Bakhshandeh, and Levin]{DoBa11}
{dos Santos},~A.~P.; Bakhshandeh,~A.; Levin,~Y. Effects of the Dielectric
  Discontinuity on the Counterion Distribution in a Colloidal Suspension.
  \emph{J. Chem. Phys.} \textbf{2011}, \emph{135}, 044124--144128\relax
\mciteBstWouldAddEndPuncttrue
\mciteSetBstMidEndSepPunct{\mcitedefaultmidpunct}
{\mcitedefaultendpunct}{\mcitedefaultseppunct}\relax
\EndOfBibitem
\bibitem[Diehl et~al.(2012)Diehl, {dos Santos}, and Levin]{DiDo12}
Diehl,~A.; {dos Santos},~A.~P.; Levin,~Y. Surface Tension of an Electrolyte-Air
  Interface: a Monte Carlo Study. \emph{J. Phys.: Condens. Matter}
  \textbf{2012}, \emph{24}, 284115--284119\relax
\mciteBstWouldAddEndPuncttrue
\mciteSetBstMidEndSepPunct{\mcitedefaultmidpunct}
{\mcitedefaultendpunct}{\mcitedefaultseppunct}\relax
\EndOfBibitem
\bibitem[Lue and Linse(2011)Lue, and Linse]{LuLi11}
Lue,~L.; Linse,~P. Macroion Solutions in the Cell Model Studied by Field Theory
  and Monte Carlo Simulations. \emph{J. Chem. Phys.} \textbf{2011}, \emph{135},
  224508--224517\relax
\mciteBstWouldAddEndPuncttrue
\mciteSetBstMidEndSepPunct{\mcitedefaultmidpunct}
{\mcitedefaultendpunct}{\mcitedefaultseppunct}\relax
\EndOfBibitem
\bibitem[Wang and Wang(2013)Wang, and Wang]{WaWa13}
Wang,~R.; Wang,~Z.~G. Effects of Image Charges on Double Layer Structure and
  Forces. \emph{J. Chem. Phys.} \textbf{2013}, \emph{139}, 124702--124109\relax
\mciteBstWouldAddEndPuncttrue
\mciteSetBstMidEndSepPunct{\mcitedefaultmidpunct}
{\mcitedefaultendpunct}{\mcitedefaultseppunct}\relax
\EndOfBibitem
\bibitem[Jing et~al.(2015)Jing, Jadhao, Zwanikken, and {de la Cruz}]{JiJa15}
Jing,~Y.~F.; Jadhao,~V.; Zwanikken,~J.~W.; {de la Cruz},~M.~O. Ionic Structure
  in Liquids Confined by Dielectric Interfaces. \emph{J. Chem. Phys.}
  \textbf{2015}, \emph{143}, 194508--194522\relax
\mciteBstWouldAddEndPuncttrue
\mciteSetBstMidEndSepPunct{\mcitedefaultmidpunct}
{\mcitedefaultendpunct}{\mcitedefaultseppunct}\relax
\EndOfBibitem
\bibitem[{dos Santos} and Levin(2015){dos Santos}, and Levin]{DoLe15}
{dos Santos},~A.~P.; Levin,~Y. Electrolytes between Dielectric Charged
  Surfaces: Simulations and Theory. \emph{J. Chem. Phys.} \textbf{2015},
  \emph{142}, 194104--194110\relax
\mciteBstWouldAddEndPuncttrue
\mciteSetBstMidEndSepPunct{\mcitedefaultmidpunct}
{\mcitedefaultendpunct}{\mcitedefaultseppunct}\relax
\EndOfBibitem
\bibitem[Pianegonda et~al.(2005)Pianegonda, Barbosa, and Levin]{PiBa05}
Pianegonda,~S.; Barbosa,~M.~C.; Levin,~Y. Charge Reversal of Colloidal
  Particles. \emph{Europhys. Lett.} \textbf{2005}, \emph{71}, 831--837\relax
\mciteBstWouldAddEndPuncttrue
\mciteSetBstMidEndSepPunct{\mcitedefaultmidpunct}
{\mcitedefaultendpunct}{\mcitedefaultseppunct}\relax
\EndOfBibitem
\bibitem[Shklovskii(1999)]{Sh99}
Shklovskii,~B.~I. Screening of a Macroion by Multivalent Ions:
  Correlation-Induced Inversion of Charge. \emph{Phys. Rev. E} \textbf{1999},
  \emph{60}, 5802--5811\relax
\mciteBstWouldAddEndPuncttrue
\mciteSetBstMidEndSepPunct{\mcitedefaultmidpunct}
{\mcitedefaultendpunct}{\mcitedefaultseppunct}\relax
\EndOfBibitem
\bibitem[{dos Santos} et~al.(2009){dos Santos}, Diehl, and Levin]{DoDi09}
{dos Santos},~A.~P.; Diehl,~A.; Levin,~Y. Electrostatic Correlations in
  Colloidal Suspensions: Density Profiles and Effective Charges beyond the
  Poisson-Boltzmann Theory. \emph{J. Chem. Phys.} \textbf{2009}, \emph{130},
  124110--214113\relax
\mciteBstWouldAddEndPuncttrue
\mciteSetBstMidEndSepPunct{\mcitedefaultmidpunct}
{\mcitedefaultendpunct}{\mcitedefaultseppunct}\relax
\EndOfBibitem
\bibitem[Levin(2002)]{Le02}
Levin,~Y. Electrostatic Correlations: from Plasma to Biology. \emph{Rep. Prog.
  Phys.} \textbf{2002}, \emph{65}, 1577--1632\relax
\mciteBstWouldAddEndPuncttrue
\mciteSetBstMidEndSepPunct{\mcitedefaultmidpunct}
{\mcitedefaultendpunct}{\mcitedefaultseppunct}\relax
\EndOfBibitem
\bibitem[{Quesada-P\'erez} et~al.(2002){Quesada-P\'erez},
  {Callejas-Fern\'andez}, and {Hidalgo-\'Alvarez}]{QuCa02}
{Quesada-P\'erez},~M.; {Callejas-Fern\'andez},~J.; {Hidalgo-\'Alvarez},~R.
  Interaction Potentials, Structural Ordering and Effective Charges in
  Dispersions of Charged Colloidal Particles. \emph{Adv. Colloid Interface
  Sci.} \textbf{2002}, \emph{95}, 295--315\relax
\mciteBstWouldAddEndPuncttrue
\mciteSetBstMidEndSepPunct{\mcitedefaultmidpunct}
{\mcitedefaultendpunct}{\mcitedefaultseppunct}\relax
\EndOfBibitem
\bibitem[{Fern\'andez-Nieves} et~al.(2005){Fern\'andez-Nieves},
  {Fern\'andez-Barbero}, {de las Nieves}, and Vincent]{FeFe05}
{Fern\'andez-Nieves},~A.; {Fern\'andez-Barbero},~A.; {de las Nieves},~F.~J.;
  Vincent,~B. Ionic Correlations in Highly Charge-Asymmetric Colloidal Liquids.
  \emph{J. Chem. Phys.} \textbf{2005}, \emph{123}, 054905--054908\relax
\mciteBstWouldAddEndPuncttrue
\mciteSetBstMidEndSepPunct{\mcitedefaultmidpunct}
{\mcitedefaultendpunct}{\mcitedefaultseppunct}\relax
\EndOfBibitem
\bibitem[Solis and {de la Cruz}(2001)Solis, and {de la Cruz}]{SoDe01}
Solis,~F.~J.; {de la Cruz},~M.~O. Flexible Polymers Also Counterattract.
  \emph{Phys. Today} \textbf{2001}, \emph{54}, 71--72\relax
\mciteBstWouldAddEndPuncttrue
\mciteSetBstMidEndSepPunct{\mcitedefaultmidpunct}
{\mcitedefaultendpunct}{\mcitedefaultseppunct}\relax
\EndOfBibitem
\end{mcitethebibliography}

\newpage
\section{Table of contents only}

\begin{figure}[h]
\vspace{0.2cm}
\includegraphics[width=5cm]{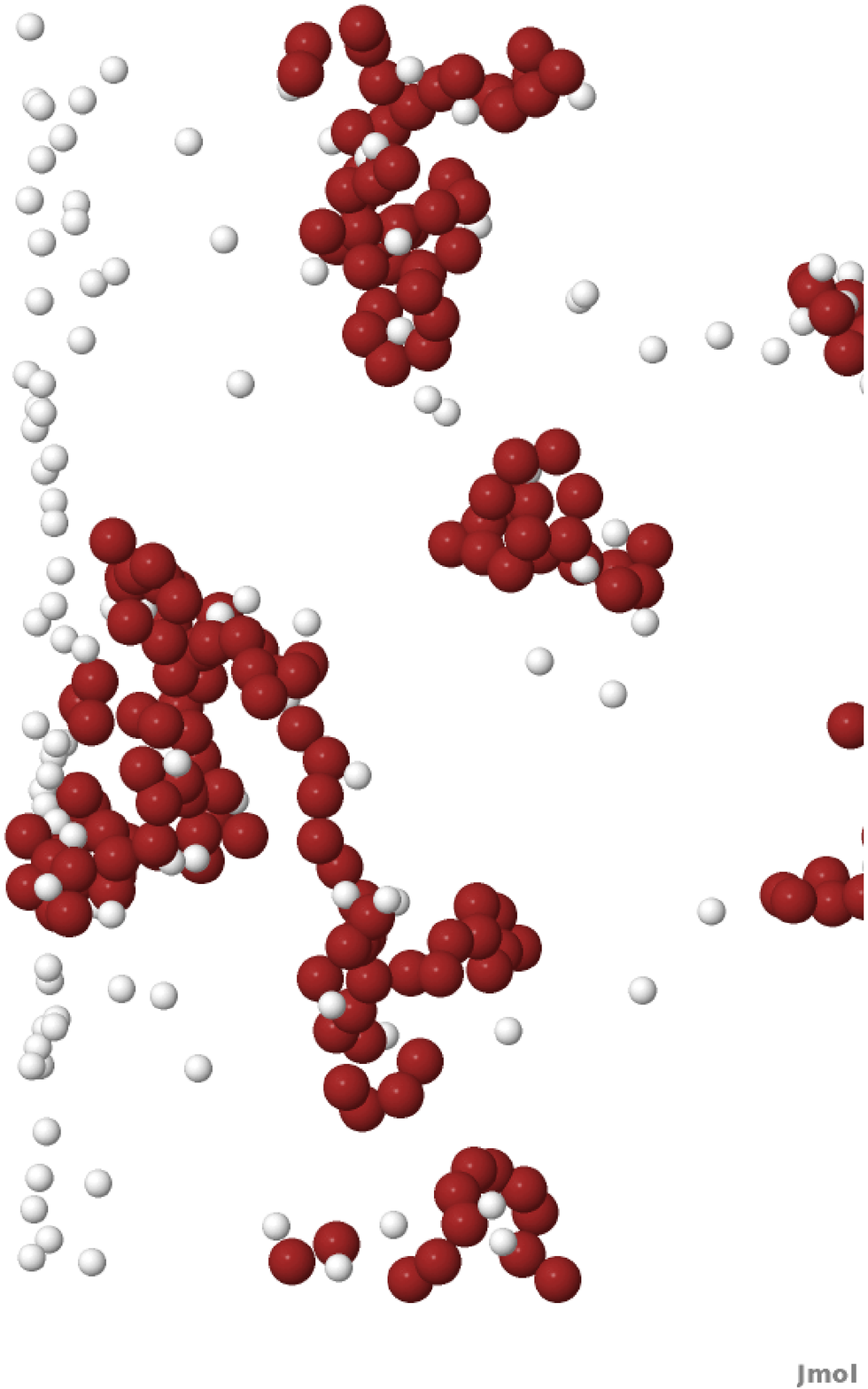}\vspace{0.2cm}
\label{toc}
\end{figure}

\end{document}